%% file: main.tex
\documentclass{Interspeech2024}

\usepackage{amsmath,amsfonts}
\usepackage{algorithm}
\usepackage{array}
\usepackage[caption=false,font=normalsize,labelfont=sf,textfont=sf]{subfig}
\usepackage{textcomp}
\usepackage{stfloats}
\usepackage{url}
\usepackage{verbatim}
\usepackage{graphicx}
\usepackage{cite}

\usepackage{amsthm,amsopn}
\usepackage{bm}
\usepackage{hyperref}
\usepackage{microtype}
\usepackage{url}
\usepackage[capitalise]{cleveref}
\usepackage{graphicx,caption,lipsum}
\usepackage{booktabs,multirow}
\usepackage{placeins}
\usepackage{algpseudocode,algorithm}
\usepackage{color}
\usepackage[table,xcdraw]{xcolor}
\definecolor{Gray}{gray}{0.9}
\definecolor{brightturquoise}{rgb}{0.85, 1, 1}
\usepackage{bbm}
\usepackage{enumerate}   
\usepackage{adjustbox}
\usepackage[normalem]{ulem}
\usepackage{fixltx2e}
\usepackage{wrapfig}
\usepackage{threeparttable}
\usepackage{tikz}
\usepackage{booktabs}

\usepackage{seqsplit}

\interspeechcameraready 

\title{YOLO-Stutter: End-to-end Region-Wise Speech Dysfluency Detection}

\name[affiliation={1}]{Xuanru}{Zhou}
\name[affiliation={2}]{Anshul}{Kashyap}
\name[affiliation={3}]{Steve}{Li}
\name[affiliation={2}]{Ayati}{Sharma}
\name[affiliation={4}]{Brittany}{Morin}
\name[affiliation={4}]{David}{Baquirin}
\name[affiliation={4}]{Jet}{Vonk}
\name[affiliation={4}]{Zoe}{Ezzes}
\name[affiliation={4}]{Zachary}{Miller}
\name[affiliation={4}]{Maria Luisa}{Gorno-Tempini}
\name[affiliation={2 \dagger}]{Jiachen}{Lian}
\name[affiliation={2 \dagger}]{Gopala}{Krishna Anumanchipalli}

\address{
   $^1$ Zhejiang University, China \quad $^2$ University of California, Berkeley \\
   $^3$ Harvard University \quad $^4$ University of California, San Francisco 
}
  
\email{xuanruzhou15@gmail.com, \{jiachenlian, gopala\}@berkeley.edu}


\begin{document}
\maketitle
\input{abstract}

\input{introduction}

\input{simulation}

\input{detection}
\input{experiments}
\input{conclusion}

\bibliographystyle{IEEEtran}
\bibliography{mybib}

\end{document}

%% file: abstract.tex
\begin{abstract}
Dysfluent speech detection is the bottleneck for disordered speech analysis and spoken language learning. Current state-of-the-art models are governed by rule-based systems~\cite{lian2023unconstrained-udm, lian2024hierarchical} which lack efficiency and robustness, and are sensitive to template design. In this paper, we propose \textit{YOLO-Stutter}: a \textit{first end-to-end} method that detects dysfluencies in a time-accurate manner. YOLO-Stutter takes  \textit{imperfect speech-text alignment} as input, followed by a spatial feature aggregator, and a temporal dependency extractor to perform region-wise boundary and class predictions. We also introduce two dysfluency corpus, \textit{VCTK-Stutter} and \textit{VCTK-TTS}, that simulate natural spoken dysfluencies including repetition, block, missing, replacement, and prolongation. Our end-to-end method achieves \textit{state-of-the-art performance} with a \textit{minimum number of trainable parameters} for on both simulated data and real aphasia speech . Code and datasets are open-sourced at \url{https://github.com/rorizzz/YOLO-Stutter}
\end{abstract}

\begin{figure*}[ht]
    \centering
    \includegraphics[height=7.2cm]{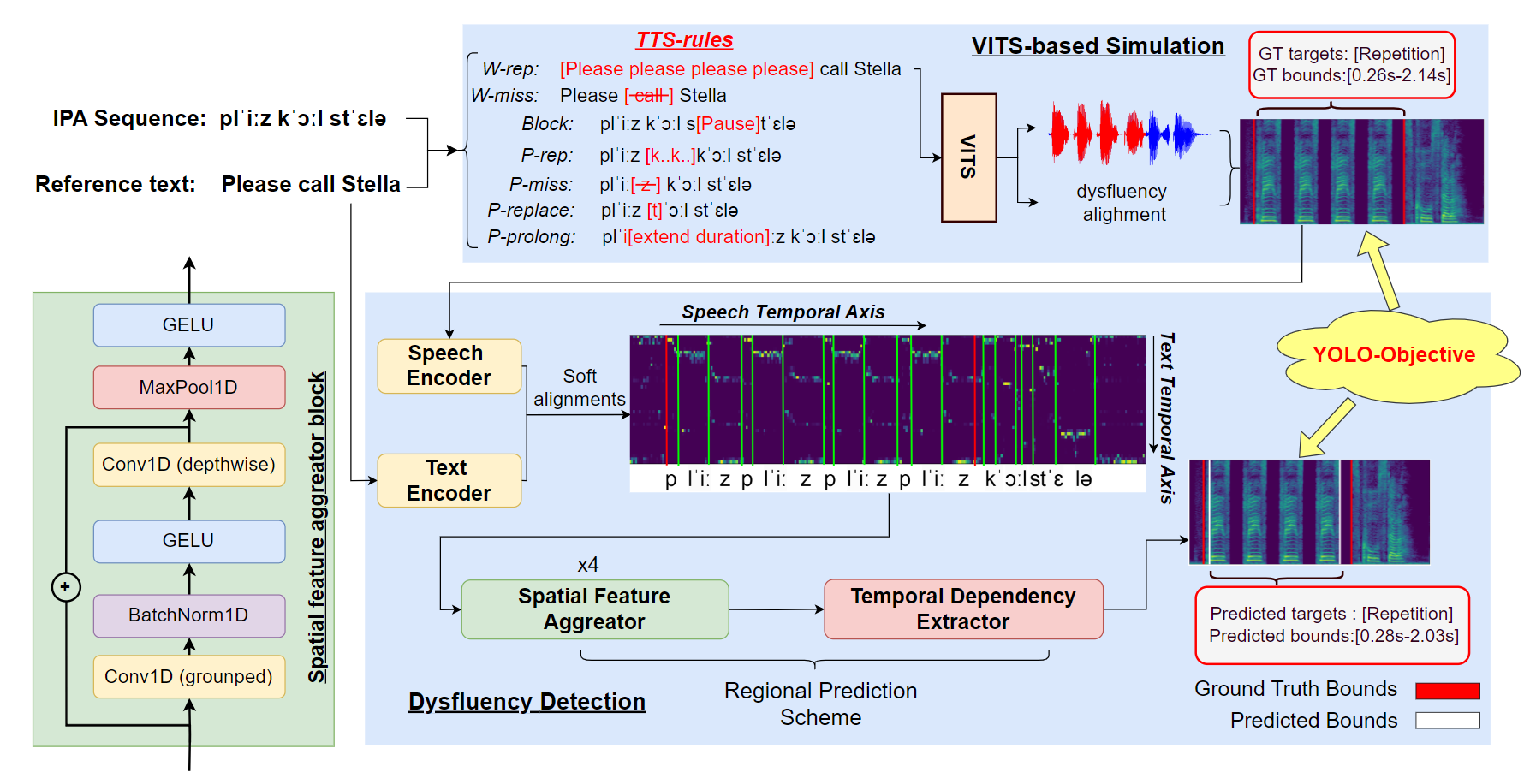}
    \caption{YOLO-Stutter method's workflow: starting from reference text and its IPA sequence, by applying TTS-rules and VITS model, we generate the dysfluent speech along with its dysfluent alignment. Utilizing pretrained VITS speech and text encoders, we produce a soft-alignment matrix from the given reference text and mel spectrogram of dysfluent speech. Subsequently, the matrix is then processed through spatial feature aggreator and temporal dependency extractor, leading to predicted targets and bounds. The architecture of our spatial feature aggreator block is illustrated in the left green block. Here are examples of our simulated speech
    \url{https://bit.ly/3PkKE8W}}
    \label{yolo-stutter}
\end{figure*}

%% file: introduction.tex
\section{Introduction}

Speech dysfluencies are defined as any form of speech that contains sound repetition, deletion, insertion, replacement, prolongation, block, etc.\cite{lian2023unconstrained-udm, lian2024hierarchical, palfy2012pattern-dysfluent, kouzelis2023weakly}. Dysfluency modeling holds significant promise in disordered speech screening\cite{nidcd}, and has a huge impact on the language learning market~\cite{vcl}. 
Dysfluency modeling is a speech transcription problem, which seems to be well tackled by recent large-scale developments~\cite{radford2022whisper,lian2023av-data2vec, zhang2023google-usm, pratap2023scaling-speech}. However, large ASR models struggle with dysfluent speech for two reasons: (1) they transcribe speech into words or phonemes that lack dysfluency-related information, and (2) their strong language models often interpret dysfluent speech as fluent, e.g., transcribing "P-P-Please call Stella" as "Please call Stella" or "Praise call Stella". Progress in dysfluency modeling has been limited due to the lack of a well-defined problem formulation and high-quality annotated dysfluency data.

UDM~\cite{lian2023unconstrained-udm} first formally defines \textit{dysfluency modeling} as the detection of all types of dysfluencies at both word and phoneme levels, with accurate time region prediction. UDM transcribes dysfluent speech into unconstrained forced alignment and adopts \textit{2D-Alignment} for detection task. H-UDM~\cite{lian2024hierarchical} developed a recursive 2D alignment algorithm that further boosts performance. These methods represent the de-facto pipeline at the time. 

In this work, we tackle dysfluency modeling from a totally different perspective. Still following the dysfluency modeling criterion~\cite{lian2023unconstrained-udm}, we develop an \textit{end-to-end model} which directly predicts dysfluencies and time regions from dysfluent speech and reference text input without any handcraft templates. To create training data, we introduce \textit{VCTK-TTS} (7X larger than VCTK~\cite{yamagishi2019cstr-vctk}), a synthetic dysfluency dataset created using the VITS~\cite{kim2021conditional-vits}, including repetition, missing, block, replacement, and prolongation at both the phoneme \& word levels. VCTK-TTS offers a more natural representation of speech dysfluencies compared to VCTK++~\cite{lian2023unconstrained-udm}, and the creation process is automated. In addition, we extend VCTK++ by incorporating word-level dysfluency and obtain a new dataset named \textit{VCTK-Stutter} (5X larger than VCTK), thus achieving word-phoneme detection. Our newly proposed datasets have the potential to set a standard benchmark for studies in this field. For the dysfluency detection task, we drew inspiration from the YOLO ~\cite{redmon2016look} and devised a region-wise prediction scheme that captures both spatial and temporal information. We developed \textit{YOLO-Stutter}, which takes soft speech-text alignments~\cite{kim2021conditional-vits} as input, followed by a spatial feature aggregator and a temporal dependency extractor to directly predict dysfluency types and time regions. We also collaborated with clinical partners and obtained data from 38 Aphasia subjects. Results on simulated speech, public corpora, and Aphasia speech indicate that YOLO-Stutter achieves state-of-the-art performance even with a minimum number of trainable parameters.

%% file: simulation.tex
\section{Naturalistic Dysfluency Simulation}
End-to-end modeling of speech dysfluency requires high-quality annotations. The prevalent approach is through simulation. Currently, existing simulated datasets~\cite{kourkounakis2021fluentnet, harvill2022frame-level-stutter, lian2023unconstrained-udm} either do not incorporate time steps and broad coverage of dysfluencies or are not naturalistic. To develop a comprehensive and naturalistic simulated dysfluency corpus, we have taken two steps: 
\begin{itemize}
    \item We extended VCTK++ ~\cite{lian2023unconstrained-udm} by introducing rule-based word-level dysfluencies in \textit{acoustic space}, creating a new dataset named \textbf{VCTK-stutter}~\cite{yamagishi2019cstr-vctk}). 
    \item We designed \textit{TTS rules} and injected phoneme and word dysfluencies in \textit{text space}. VITS~\cite{kim2021conditional-vits} was used to generate naturalistic dysfluent speech. This dataset is named \textbf{VCTK-TTS}. 
\end{itemize}

\subsection{VCTK-Stutter}
VCTK++\cite{lian2023unconstrained-udm} was created in the acoustic space by editing audio based on forced alignment. However, it only focuses on phoneme-level dysfluencies. We extended VCTK++ by incorporating word-level dysfluencies (named \textit{VCTK-Stutter}). Similar to obtaining phoneme-level alignments through MFA~\cite{mcauliffe2017montrealmfa}, in VCTK++ we employ WhisperX \cite{bain2023whisperx} to get word-level alignments, enabling us to establish precise insertion points for dysfluencies. Utilizing the word alignments generated by WhisperX, we annotate specific regions where artificial dysfluencies are to be introduced with the type of word-level dysfluency and the start and end times. We enable each word to have an equal probability of experiencing a dysfluency for consistency. The word-level dysfluency types we incorporate include: 
\begin{itemize}
    \item  \textbf{Word-repetition}: This involves the repetition of entire words or phrases multiple times, with the count of repetitions varying randomly from 1 to 4, appended by a sample of silence of length 70\% of the original word after each repetition.
    \item \textbf{Word-missing}: This pertains to the omission of certain words. The randomly chosen missing word is replaced with a silence of equivalent duration
\end{itemize}

\subsection{VCTK-TTS}
Traditional rule-based simulation methods~\cite{kourkounakis2021fluentnet, lian2023unconstrained-udm, harvill2022frame-level-stutter} operate in acoustic space, and the generated samples are not naturalistic. We developed a new pipeline that simulates in text space. To achieve this, we first convert a sentence into an IPA phoneme sequence. Then, we develop TTS rules for phoneme editing to simulate dysfluency. These rules are applied to the entire VCTK dataset~\cite{kim2021conditional-vits}, allowing the voice of generated speech to vary from the 109 speakers included in the VCTK, thus enhancing the scalability of the dataset. We call this \textbf{VITS-based Simulation}. The entire pipeline is detailed in Sec.\ref{method-pipleline}, and TTS rules are discussed in Sec.\ref{tts-rules}.
Statistics are listed in Table. 
\ref{statistics}. Overall VCTK-Stutter/VCTK-TTS are 5X/7X larger than VCTK. 

\begin{table}[h]
    \caption{Types of Dysfluency Data in VCTK-Stutter \& VCTK-TTS }
    \label{dysfluency-stats-vctk-tts}
    \centering
    \setlength{\tabcolsep}{10pt} 
    \renewcommand{\arraystretch}{1.1} 
    \resizebox{8cm}{!}{
    \begin{tabular}{l c c c c} 
    \toprule
    Dysfluency & \# Samples & Percentage &
    \# Samples  & Percentage \\
    \phantom{''} & VCTK-Stutter & VCTK-Stutter &
    VCTK-TTS  & VCTK-TTS \\
    \hline
    \hline
    Prolongation & 43738 & 19.94 & 41084 & 13.37 \\
    Block & 43959 & 20.040 & 49365 & 16.06 \\
    Replacement & 0 & 0 & 46320 & 15.07 \\
    Repetition (Phoneme) & 43738 & 19.94 & 40940 & 13.32 \\
    Repetition (Word) & 43959 & 20.04 & 41800 & 13.60 \\
    Missing (Phoneme) & 0 & 0 & 46320 & 15.07 \\
    Missing (Word) & 43959 & 20.04 & 41559 & 13.52 \\ 
    \midrule
    Total Hours of Audio & 304.66 & \phantom{''}  & 426.93 & \phantom{''}  \\
    \bottomrule
    \end{tabular}}
    \label{statistics}
\end{table}
\vspace{-7pt}
\subsubsection{Method pipeline} \label{method-pipleline}
VCTK-TTS pipelines can be divided into following steps:
\textbf{(i) Dysfluency injection:} We first convert ground truth reference text of VCTK text~\cite{yamagishi2019cstr-vctk} into IPA sequences via the VITS phonemizer~\cite{kim2021conditional-vits}, then add different types of dysfluencies at the phoneme level according to the \textit{TTS rules}.
\textbf{(ii) VITS inference: } We take dysfluency-injected IPA sequences as inputs, conduct the VITS inference procedure and obtain the dysfluent speech. 
\textbf{(iii) Annotation:} We retrieve phoneme alignments from VITS duration model, annotate the type of dysfluency on the dysfluent region.

\subsubsection{TTS rules} \label{tts-rules}
We inject dysfluency in the text space following these rules:

\begin{itemize}

\item \textbf{Repetition} (phoneme\&word): The first phoneme or syllable of a randomly picked word was repeated 2-4 times, with pauselengths varying between 0.5 to 2.0 seconds.
\item \textbf{Missing} (phoneme\&word): 
We simulated two phonological processes that characterize disordered speech\cite{Bauman-Wängler_2020} - weak syllable deletion (deletion of a random unstressed syllable based on stress markers\footnote{https://github.com/timmahrt/pysle}) and final consonant deletion. 

\item \textbf{Block}: A duration of silence between 0.5-2.0 seconds was inserted after a randomly chosen word in between the sentence.

\item \textbf{Replacement} (phoneme): We simulated fronting, stopping, gliding, deaffrication - processes that characterize disordered speech\cite{Bernthal_Bankson_Flipsen_2017} - by replacing a random phoneme with one that would mimic the phonological processes mentioned above. 

\item \textbf{Prolongation} (phoneme): 
The duration of a randomly selected phoneme in the utterance was extended by a factor randomly chosen between 10 to 15 times its original length, as determined by the duration model.

\end{itemize}

\subsection{Evaluation}
To evaluate the rationality and naturalness of two datasets we constructed, we collected Mean Opinion Score (MOS, 1-5) ratings from 10 people.  The final results are as displayed in Table. \ref{mos}.  VCTK-TTS was perceived to be far more natural than VCTK-Stutter (MOS of 4.13 compared to 2.22). We employ VCTK-Stutter as a baseline corpus.
\begin{table}[h]
    \caption{MOS for VCTK-Stutter \& VCTK-TTS Samples}
    \label{dysfluency-stats-vctk-tts}
    \centering
    \setlength{\tabcolsep}{13pt} 
    \renewcommand{\arraystretch}{1.2} 
    \resizebox{8cm}{!}{
    \small
    \begin{tabular}{l c c} 
    \toprule
    Dysfluency Type & VCTK-Stutter MOS & VCTK-TTS MOS \\
    \hline
    \hline
    Block (word) & 2.80 ± 0.63 & 3.40 ± 0.52 \\
    Missing (phoneme) & N/A & 4.70 ± 0.48 \\
    Missing (word) & 2.90 ± 0.32 & 4.80 ± 0.63 \\
    Prolong (phoneme) & N/A & 3.80 ± 0.92 \\
    Repetition (phoneme) & 1.40 ± 0.70 & 4.60 ± 0.84 \\
    Repetition (word) & 2.80 ± 0.79 & 3.70 ± 1.16 \\
    Replacement (phoneme) & N/A & 3.90 ± 0.99 \\
    \midrule
    Overall & 2.22 ± 0.84 & 4.13 ± 0.56
    \\
    \bottomrule
    \end{tabular}}
    \label{mos}
\end{table}
\vspace{-13pt}

%% file: detection.tex
\section{End-to-End Dysfluency Detection} \label{sec:e2edetection}

\begin{table*}[h]
    \caption{Accuracy (Acc) and Bound loss (BL) of the five dysfluency types trained on the VCTK-Stutter and VCTK-TTS.}
    \label{dysfluency-eval-train}
    \centering
    \setlength{\tabcolsep}{8pt} 
    \renewcommand{\arraystretch}{1.1} 
    \resizebox{15cm}{!}{
    \small
    \begin{tabular}{l c l|c c| c c| c c |c c |c c} 
     \toprule
     & Trainable & & \multicolumn{2}{c|}{Rep}& \multicolumn{2}{c|}{Block} & \multicolumn{2}{c|}{Miss} & \multicolumn{2}{c|}{Replace} & \multicolumn{2}{c}{Prolong}\\
    Methods& parameters & Dataset& Acc.\% & BL & Acc.\%  & BL & Acc.\%  & BL & Acc.\%  & BL & Acc.\%  & BL \\
    \hline
    \hline
    H-UDM~\cite{lian2024hierarchical} & 92M & VCTK-Stutter Testset & 82.08 &30ms &97.12 &27ms&20 &24ms&-&-&-&-\\ 
    YOLO-Stutter(VCTK-Stutter) & 33M & VCTK-Stutter Testset & \textbf{99.16} & \textbf{26ms}& 99.29 & 25ms & \textbf{80.00} & 18ms & - & - & \textbf{91.84} & 35ms \\
    YOLO-Stutter(VCTK-TTS)  & 33M & VCTK-Stutter Testset & 83.11 & 27ms & \textbf{100} & \textbf{22ms} & 40.00 & \textbf{17ms} & - & - & 90.34 & \textbf{34ms} \\
    \midrule
    H-UDM~\cite{lian2024hierarchical} & 92M & VCTK-TTS Testset & 74.66 &68ms & 88.44& 85ms&15&100ms &-&-&-&-\\
    YOLO-Stutter(VCTK-Stutter) & 33M &VCTK-TTS Testset & 78.31 & 66ms & 92.44 &  \textbf{43ms}& 43.33 &  42ms& - & - & 88.17 & 42ms\\
    YOLO-Stutter(VCTK-TTS) & 33M &VCTK-TTS Testset & \textbf{98.78} & \textbf{27ms} & \textbf{98.71} & 78ms & \textbf{70.00} &  \textbf{8ms} & \textbf{73.33} & \textbf{10ms} &  \textbf{93.74} & \textbf{32ms}\\
    \bottomrule
    \end{tabular}}
\end{table*}

Dysfluency modeling is text-dependent~\cite{lian2023unconstrained-udm}. We adopt the \textit{soft speech-text alignment} from VITS~\cite{kim2021conditional-vits} as input, and adopt a YOLO-like objective for the detection.  
We take 1D extension of 2D object detection from~\cite{redmon2016look}, which utilizes a region-wise prediction scheme designed to aggregate local spatial information, to develop our detector. 
The entire paradigm is shown in Fig. \ref{yolo-stutter} and the corresponding modules are detailed in the following. 

\subsection{Soft speech-text alignments}
\label{ssec:vits-alignments}

VITS~\cite{kim2021conditional-vits} takes in speech and text data to train alignment, producing a $|c_{text}| \times |z|$ monotonic attention matrix $A$ that represents how each input phoneme expands to be time-aligned with target speech, where $c_{text}$ represents the tokenized text dimension and $z$ the temporal speech dimension. We use the soft alignments $A$ and apply a softmax operation across the text dimension, computing the maximum attention value for each time step. We use pre-trained text and speech encoders from~\cite{kim2021conditional-vits}.

\subsection{Spatial feature aggregator}
\label{ssec:subhead}

To preserve local spatial features for region-wise predictions, we use learnable spatial feature aggregator blocks. Unlike spectrograms, soft alignment data cannot be effectively processed using traditional conformer~\cite{gulati2020conformer} methods with pointwise and depthwise convolutions, as collapsing information across either the text or speech dimension would severely corrupt the relevant signal. Our initial experiments confirm that this design fails to learn dysfluencies. To address this, we employ a series of depthwise and grouped convolution operations, enabling iterative downsampling while preserving regional spatial features.

\subsection{Temporal dependency extractor}
\label{ssec:subhead}

We found that using only spatial feature aggregator modules is insufficient for accurately detecting dysfluencies due to their reliance on local spatial information. To effectively capture the sequential nature of dysfluent speech, we employ a second-stage Transformer encoder~\cite{vaswani2023attenion} that treats each speech region as an element in a sequence, using the text dimension as the embedding dimension. This Transformer encoder consists of 8 layers with 8 attention heads per layer, enabling the extraction of longer-term temporal information crucial for dysfluency detection.

\subsection{YOLO objective}
\label{yolo-objective}
Following YOLO\cite{redmon2016look}, we use a weighted loss-based multi-task training pipeline with three separate loss values: dysfluency presence confidence loss (binary cross-entropy), start and end bound loss (mean squared error), and dysfluency type categorical loss (cross-entropy). Confidence loss is computed across all regions, categorical loss in regions with dysfluency, and bound loss on the region ``responsible" for the dysfluency. Bound values are normalized between 0-1 using fixed padded lengths as max bound values. Loss is shown below:
\vspace{-8pt}

\parbox{23em}{
\begin{flushright}
$\mathbb{L}=\lambda_{\text{bound}} \frac{1}{S} \displaystyle\sum_{i=0}^{S} \mathbbm{1}_{\text{obj}} [(b_{\text{start}} - \hat{b}_{\text{start}})^2 + (b_{\text{end}} - \hat{b}_{\text{end}})^2]$ \\[-1ex]
$-\lambda_{\text{conf}}  \frac{1}{S} \displaystyle\sum_{i=0}^{S}  \hat{y}_i \log(p(y_i)) + (1 - \hat{y}_i) \cdot \log(1 - p(y_i))$\\[-1ex]
$-\lambda_{\text{class}} \frac{1}{S} \displaystyle\sum_{i=0}^{S} \displaystyle\sum_{j=0}^{n} c_n \log (p(\hat{c}_n))$ 
\end{flushright}
}
where $\mathbbm{1}_{obj}$ denotes the presence of a dysfluency in a specific region, $y_i$ \& $\hat{y}_i$ denote the predicted \& target confidence values, and $c$ \& $\hat{c}$ denote the predictions \& targets for n classes. Loss is computed and averaged across S regions within a single sample. We scale the bound loss term by a factor of 5 and the classification term by a factor of 0.5. 

%% file: experiments.tex
\section{Experiments}


\subsection{Datasets}

\textbf{(1) VCTK~\cite{yamagishi2019cstr-vctk} }includes 109 native English speakers with accented speech.  It is used in both \textit{VCTK-Stutter} and \textit{VCTK-TTS}. \textbf{(2) LibriStutter~\cite{kourkounakis2021fluentnet}}  contains artificially stuttered speech and stutter classification labels for 5 stutter types. It was generated using 20 hours of audio selected from \cite{panayotov2015librispeech}. \textbf{(3) Aphasia Speech} is collected from our clinical collaborators, our dysfluent data comprises 38 participants diagnosed with Primary Progressive Aphasia (PPA), larger than the data used in~\cite{lian2023unconstrained-udm, lian2024hierarchical} which only has 3 speakers. People were asked to read grandfather passage, leading about 1 hour of speech. \textbf{(4) UCLASS~\cite{Howell2009TheUA}}
 contains recordings from 128 children and adults who stutter. Only 25 files have been annotated and did not annotate for the block class, we only used those files and did not use the block class for subsequent datasets. \textbf{(5) SEP-28K} is curated by~\cite{lea2021sep}, contains 28,177 clips extracted from publicly available podcasts. We removed files where the annotations were labelled as ``unsure".

\begin{table}[h]
    \caption{Type-specific accuracy (ACC) and time F1-score}
    \label{compare-with-benchmark}
    \centering
    \setlength{\tabcolsep}{5pt} 
    \renewcommand{\arraystretch}{1} 
    \resizebox{8cm}{!}{
    \small
    \begin{tabular}{l c| c c c| c} 
     \toprule
    Methods & Dataset & \multicolumn{3}{c|}{Accuracy ($\%$, $\uparrow$)} & Time F1 ($\uparrow$)\\
    \rowcolor{lightgray}
    & & \textit{Rep} & \textit{Prolong} & \textit{Block} &\\ 
    \midrule
    Kourkounakis et al. ~\cite{kourkounakis2021fluentnet}& UCLASS & 84.46 & 94.89 & - & 0\\
    Jouaiti et al. ~\cite{segment-detection4} & UCLASS & 89.60 & 99.40 & - & 0\\
    Lian et al. ~\cite{lian2024hierarchical} & UCLASS & 75.18&- & 50.09 & 0.700\\
    YOLO-Stutter (VCTK-Stutter) & UCLASS &76.82 & 78.49 & 54.13 & 0.799\\
    \textbf{YOLO-Stutter (VCTK-TTS)} & UCLASS & 92.00 & 91.43 & 56.00  & \textbf{0.893}\\
    \midrule
    Kourkounakis et al. ~\cite{kourkounakis2021fluentnet}& LibriStutter & 82.24& 92.14 & - & 0\\
    Lian et al. ~\cite{lian2024hierarchical} &LibriStutter &85.00& -& -& 0.660\\
    YOLO-Stutter (VCTK-Stutter) & LibriStutter &88.26 & 65.32 & - & 0.684\\
    \textbf{YOLO-Stutter (VCTK-TTS)} & LibriStutter& 89.71 & 67.74& - & \textbf{0.697}\\
    \midrule
   Jouaiti et al. ~\cite{segment-detection4} &SEP-28K& 78.70& 93.00 & - & 0 \\
    Lian et al. ~\cite{lian2024hierarchical} &SEP-28K & 70.99& -&66.44&0.699 \\
    YOLO-Stutter (VCTK-Stutter) &SEP-28K& 
     69.79 &  63.72 & 71.70 & 0.747\\
    \textbf{YOLO-Stutter (VCTK-TTS)} & SEP-28K& 82.01&  89.19& 68.09 & \textbf{0.813}\\    
    \bottomrule
    \end{tabular}}
\end{table}

\begin{table}[htp!]
    \caption{Dysfluency evaluation on Aphasia speech.}
    \label{dysfluency-eval-ppa}
    \centering
    \setlength{\tabcolsep}{8pt} 
    \renewcommand{\arraystretch}{1.2} 
    \resizebox{8cm}{!}{
    \small
    \begin{tabular}{l c c c} 
    \toprule
    Methods & Ave. Acc. ($\%$, $\uparrow$) & Best Acc.($\%$, $\uparrow$) & Ave. BL (ms, $\downarrow$)\\
    \hline
    \hline
    \vspace{0.5mm}
    H-UDM~\cite{lian2024hierarchical} &41.8 & 70.22&52ms\\
    YOLO-Stutter(VCTK-Stutter) & 45.71 & 73.08(Rep) & 46ms\\
    YOLO-Stutter(VCTK-TTS) & 54.19 & \textbf{92.54 (Block)} & \textbf{21ms}\\
    \bottomrule
    \end{tabular}}
\end{table}

\vspace{-10pt}

\subsection{Training}
The detector is trained with a randomized 90/10 train/test split on both VCTK-Stutter and VCTK-TTS, with a batch size of 64. We utilize the Adam optimization method with beta values (0.9, 0.999) and learning rate of 3e-4, opting not to use dropout or weight-decay in our training process. For VCTK-Stutter, the model is trained for 20 epochs and total of 39 hours on a RTX A6000. Meanwhile, for VCTK-TTS, the model is trained for 30 epochs and total of 70 hours on the RTX A6000.

\subsection{Evaluation Metrics}
\textbf{(1) Accuracy} is the accuracy of correct predictions of stutter types in regions that contain a dysfluency. The confidence prediction accuracy is defined as the number of correct predictions of dysfluent and fluent regions over all the regions.
\textbf{(2) Bound loss} is the mean squared loss between the predicted bounds and the actual bounds of the dysfluent regions, normalized to the 1024-length padded spectrogram and converted to a time scale using a known 20ms sampling frequency.
\textbf{(3) Time F1~\cite{lian2023unconstrained-udm}} measures accuracy in bounds prediction, calculating based on the intersection between predicted and actual bounds. Once there is an overlap, the sample is considered as a True Positive sample.

\subsection{Validation}
To assess the performance of trained detector, we conduct evaluations on the VCTK++ and VCTK-TTS testsets, as well as on the PPA data. The evaluation results, including type-specific detection accuracy and bound loss metrics, are presented in Table~\ref{dysfluency-eval-train} for the training datasets and in Table~\ref{dysfluency-eval-ppa} for the PPA data. To compare the performance with previous works, we also validated the model on UCLASS, Libristutter and SEP-28K, calculating type-specific accuracy as well as Time F1.

\subsection{Results and Discussions}
We conducted inference on our proposed VCTK-Stutter and VCTK-TTS test sets, using H-UDM as baseline. As indicated in Table \ref{dysfluency-eval-train}, both YOLO-Stutter (VCTK-Stutter) and YOLO-Stutter (VCTK-TTS) surpassed H-UDM across all metrics. Notably, the results on VCTK-TTS from YOLO-Stutter (VCTK-TTS) were particularly promising. VCTK-TTS closely mimics human speech, affirming the potential of YOLO-Stutter (VCTK-TTS) to deliver substantial performance on actual human speech. Furthermore, we presented our findings on UCLASS, LibriStutter, and SEP-28K, as shown in Table \ref{compare-with-benchmark}. Given that the test sets from the original benchmarks are private, direct accuracy comparisons may not be entirely fair. Focusing more on time-aware detection, we reported the Time F1 score for each dataset, where all baselines, except for H-UDM, scored 0. Both of our proposed methods consistently outperformed H-UDM. Lastly, we evaluated our methods on actual PPA speech, with results detailed in Table \ref{dysfluency-eval-ppa}. Both models consistently exceeded H-UDM's performance, with VCTK-TTS achieving additional gains owing to its higher naturalness, as reflected in Table \ref{mos}. However, the average accuracy remains low, highlighting the challenge of perfectly capturing the real distribution of dysfluency.

\subsection{Ablation experiments}
To investigate the impact of the proportions of different dysfluency types quantities on training results, we selected three different proportions except for average on VCTK-TTS, as follows: $P$ = [Rep, Block, Miss, Replace, Prolong], $P_1$=[0.9 : 1 : 1 : 1 : 1], $P_2$= [1 : 1 : 1.2 : 1: 1], $P_3$=[1 : 1 : 1.2 : 1.2 : 1]. Table \ref{Ablation} shows validated type-specific accuracy on VCTK-TTS inference testsets, from which we can see that although the proportions were adjusted, the type accuracy for repetition and block has remained relatively high. The accuracy for missing increases with its sample proportion, but does not improve as the proportions of other samples (repetition) decrease. Increasing missing and replace proportions improves missing type accuracy but lowers replacement accuracy.

\vspace{-6pt}
\begin{table}[htp!]
    \caption{Type-specific accuracy of different proportions}
    \label{Ablation}
    \centering
    \setlength{\tabcolsep}{10pt}
    \renewcommand{\arraystretch}{1.2} 
     \resizebox{8cm}{!}{
     \small
    \begin{tabular}{c c c c c c} 
     \toprule
    Proportions& Rep & Block & Miss & Replace & Prolong\\
     \hline
     \hline
     Average & 98.78\% & 98.71\% & 70.00\% & 73.33\% & 93.74\% \\
     $P_1$ & 96.66\% & 99.97\% & 78.33\% & 66.67\% & 87.72\%  \\
     $P_2$ & 96.67\% & 99.97\% & 60.00\% &  46.67\% & 93.33\%\\
     $P_3$  & 96.67\% & 99.97\% &  83.29\%  & 54.32\% & 90.00\%\\
    \bottomrule
    \end{tabular}}
\end{table}
\vspace{-13pt}

%% file: conclusion.tex
\section{Conclusion and Limitations}
We introduce \textit{YOLO-Stutter}, the first end-to-end dysfluency detection paradigm, achieving state-of-the-art performance with fewer trainable parameters compared to rule-based systems. We contribute two simulated dysfluent corpora, \textit{VCTK-Stutter} and \textit{VCTK-TTS}, with high-quality annotations covering sound repetition, replacement, insertion, deletion, and block. However, limitations persist. First, average accuracy on aphasia speech inference remains limited, indicating a significant gap between simulated and real dysfluency. Second, VCTK-TTS still produces robotic noise and discontinuous sounds, warranting exploration of adversarial methods to address these issues. Third, it is worth to explore simulation in articulatory space~\cite{lian2023factor, lian22bcsnmf, inversion, wu23k_interspeech, cho2024self-universal}. Lastly, dysfluency is a clinical problem and is speaker dependent. Disentangled analysis and synthesis~\cite{lian2022utts, qian2022contentvec, lian2022robust-d-dsvae, lian2022towards-c-dsvae, choi2022nansy++} should be leveraged to tackle dysfluency in speaker-free space.

\section{Acknowledgement}
Thanks for support from UC Noyce Initiative, Society of Hellman Fellows, NIH/NIDCD and the Schwab Innovation fund.